\documentclass[12pt,nohyper]{JHEP}
\usepackage{epsf}

\newcommand  {\rf} [1]{(\ref{#1})}
\newcommand  {\beq}[1]{\begin{equation}\label{#1}}
\newcommand  {\eeq}   {\end{equation}}
\newcommand  {\bea}   {\begin{eqnarray}}
\newcommand  {\eea}   {\end{eqnarray}}
\newcommand  {\e}     {\mbox{e}}
\renewcommand{\d}     {\mbox{d}}
\newcommand  {\g}     {\gamma}

\renewcommand{\l}     {\lambda}

\renewcommand{\b}     {\beta}
\renewcommand{\t}     {\tau}

\renewcommand{\d}     {\delta}
\newcommand  {\n}     {\nu}  

\newcommand  {\vev}[1]{\left\langle #1 \right\rangle}

\newcommand  {\ra}    {\right\rangle}
\newcommand  {\la}    {\left\langle}

\newcommand  {\equ}   {\!=\!}

\title{Crossing the c=1 Barrier in 2d Lorentzian Quantum Gravity}

\author{J. Ambj\o rn\thanks{Email: ambjorn@nbi.dk},  
\\ The Niels Bohr Institute, \\
Blegdamsvej 17, DK-2100 Copenhagen \O , Denmark}
\author{K.N. Anagnostopoulos\thanks{Email: konstant@physics.uch.gr}
\\Department of Physics, University of Crete,\\ 
P.O.Box 2208, GR-710 03 Heraklion, Crete, Greece}
\author{R. Loll\thanks{Email: loll@aei-potsdam.mpg.de} 
\\ Albert-Einstein-Institut,\\
Am M\"{u}hlenberg 5, D-14476 Golm, Germany}

\preprint{NBI-HE-99-26\\ AEI-1999-17}

\abstract{In an extension of earlier work we investigate the behaviour
of two-dimensional Lorentzian quantum gravity under coupling to a
conformal field theory with $c>1$. This is done by analyzing
numerically a system of eight Ising models (corresponding to $c\equ
4$) coupled to dynamically triangulated Lorentzian geometries.  It is
known that a single Ising model couples weakly to Lorentzian quantum
gravity, in the sense that the Hausdorff dimension of the ensemble of
two-geometries is two (as in pure Lorentzian quantum gravity) and the
matter behaviour is governed by the Onsager exponents.  By increasing
the amount of matter to 8 Ising models, we find that the geometry of
the combined system has undergone a phase transition. The new phase is
characterized by an anomalous scaling of spatial length relative to
proper time at large distances, 
and as a consequence the Hausdorff dimension is now
three. In spite of this qualitative change in the geometric sector,
and a very strong interaction between matter and geometry, the
critical exponents of the Ising model retain their Onsager
values. This provides evidence for the conjecture that the KPZ values
of the critical exponents in 2d Euclidean quantum gravity are entirely
due to the presence of baby universes. Lastly, we summarize the lessons
learned so far from 2d Lorentzian quantum gravity.}

\begin{document}
\section{Introduction}\label{intro}

It may come as a surprise to practitioners of two-dimensional gravity
that there is more than one way of constructing a viable quantum
theory by path-integral methods, and that there is indeed ``life
beyond Liouville gravity". The new, alternative theory of 2d quantum
gravity in question was first constructed as the continuum limit of an
exactly soluble model of dynamically triangulated two-geometries
\cite{al}, which could be interpreted as representing {\it Lorentzian}
geometries with a causal structure and a preferred time direction.  It
has recently been shown that there is a whole universality class of
such Lorentzian models, some of which are obtained by adding a
curvature term to the gravity action or by using building blocks
different from triangles in the construction of geometries \cite{fgk}.

An investigation of Lorentzian gravity coupled to Ising spins led to
the conclusion that in spite of strong fluctuations of the underlying
geometries, the critical matter behaviour in the coupled system is
governed by the Onsager exponents \cite{aal} (which one also finds for
the Ising model on a fixed, regular lattice).  This immediately raises
the following questions: If we continue to add matter to the system,
do we eventually observe a qualitative change in the behaviour of
geometry and/or matter?  Is there an analogue of the $c\equ 1$ barrier
of Liouville quantum gravity beyond which the combined gravity-matter
system degenerates?  We address these and related issues below, by
studying numerically 8 Ising models (corresponding to a $c\equ 4$
conformal field theory) coupled to Lorentzian quantum gravity.

In order to set the stage for our present investigation, let us recall
some salient features of the Lorentzian gravity model \cite{al,alnr}.
One idea behind the formulation of such a model is to take the
Lorentzian structure seriously within a path-integral approach and in
this way bridge the gap between the canonical quantization and the
(Euclidean) path-integral formulation of gravity.  The Lorentzian
aspects of the model are two-fold: compared with the Euclidean case,
the state sum is taken over a restricted class of triangulated
two-geometries, namely, those which are generated by evolving a
one-dimensional spatial slice and allow for the introduction of a
causal structure.  Secondly, the Lorentzian propagator is obtained by
a suitable analytic continuation in the coupling constant. During time
evolution, we do not permit the spatial slice to split into several
components (i.e. change its topology), because the resulting
space-time geometry would not be compatible with our discrete notion
of causality. (In a continuum picture, the local lightcone structure
associated with a Lorentzian metric must necessarily become degenerate
at such branching points.) This is exactly the situation described by
usual canonical (quantum) gravity.

In the pure gravity model, the loop-loop correlator and various
geometric properties can be calculated exactly and compared to
Euclidean 2d quantum gravity, as given by Liouville gravity or 2d
quantum gravity defined by dynamical triangulations or matrix models.
The two models turn out to be inequivalent. For example, the Hausdorff
dimension of the Lorentzian quantum geometry is $d_H =2$, indicating a
much smoother behaviour than that of the Euclidean case where
$d_{H}\equ 4$. The difference between the fractal structures of
Lorentzian and Euclidean quantum gravity can be traced to the absence
or presence of so-called baby universes.  These are outgrowths of the
geometry taking the form of branchings-over-branchings, which are
known to dominate the typical geometry contributing to the Euclidean
state sum. Such branchings and associated topology changes with
respect to the preferred spatial slicing are absent from the histories
contributing to the Lorentzian state sum.

Baby universes, i.e. discrete evolution moves resulting in spatial
topology chan\-ges may be re-introduced by hand in the Lorentzian
formulation (if one is willing to give up causality). This corresponds
to ``switching on" an additional term in the differential equation for
the propagator, in such a way that the scaling limit must be modified
in order to produce well-defined continuum physics.

A further difference between 2d Lorentzian and Euclidean gravity is
revealed by coupling them to conformal matter.  In the Euclidean case
this is governed by the famous KPZ scaling relations.  They describe
how the critical exponents of a conformal field theory change when it
is coupled to Euclidean quantum gravity, and how the entropy exponent
$\g_{str}$ for two-geometries (the so-called string susceptibility)
changes due to their coupling to the conformal matter fields.

In 2d Lorentzian gravity, the continuum limit of the quantum geometry
was found to be {\it unchanged} under coupling to a $c\equ 1/2$
conformal field theory, in the form of an Ising model at its critical
point\footnote{The
critical point of the Ising model we refer to is the critical point of
the combined Ising-gravity system. See \cite{aal} 
for details.}. The Hausdorff dimension remains equal to two, and
an appropriately rescaled distribution of spatial volumes coincides
with the distribution found in pure Lorentzian gravity. In addition,
the values of the critical exponents of the Ising model agree with
those of the Ising model on a regular lattice.  In other words,
coupling the Ising model to Lorentzian gravity does not affect the
nature of its (second-order) phase transition.  Summarizing, one may
say that the coupling between $c\equ 1/2$ conformal matter and
geometry in Lorentzian quantum gravity is weak.

To avoid a frequent misunderstanding, we must emphasize that this is
{\it not} a trivial consequence of the fact that $d_H\equ 2$ in
Lorentzian quantum gravity. Although a flat space-time implies
$d_H\equ 2$ for the Hausdorff dimension, the converse is by no means
true.  In fact, the geometry {\it does} fluctuate strongly in
Lorentzian gravity, as was demonstrated in \cite{al,aal}.  There are
other examples to illustrate that the Hausdorff dimension is only a
very rough measure of geometry.  Consider 2d {\it Euclidean} quantum
gravity coupled to conformal field theories with $c >1$: in these
models the geometry fluctuates so wildly that the two-dimensional
surfaces are torn apart and degenerate into so-called branched
polymers, which again have $d_H=2$, the same as for smooth surfaces!

An important conclusion one can draw from the results obtained in
\cite{aal} is that the strong coupling between Euclidean quantum
gravity and conformal matter is directly caused by the presence of
baby universes.  Various qualitative arguments have been put forward
in the past to support this idea, which is of course not new.
However, one never had a model which prohibited the creation of baby
universes, and which could be used to verify explicitly that the
coupling between geometry and matter in this case is weak.  The
observed weak-coupling behaviour in the Lorentzian model opens up the
intriguing possibility that one might be able to cross the $c\equ 1$
barrier in Lorentzian 2d quantum gravity coupled to matter. This is
the issue we will study numerically in the remainder of this article,
by coupling eight Ising models to Lorentzian quantum gravity,
corresponding at the critical point of the combined system to a $c\equ
4$ conformal field theory.

\section{Coupling gravity to multiple Ising spins}

In our previous work \cite{al} we have defined the two-loop function
of Lorentzian 2d gravity as the state sum
\beq{2.1}
G(\l,t) = \sum_{T \in {\cal T}_t} e^{-\l N_T},
\eeq
where the summation is over all triangulations $T$ of cylindrical
topology with $t$ time-slices, $N_T$ counts the number of triangles in
the triangulation $T$, and $\l$ is the bare cosmological constant.
Since we are primarily interested in the bulk behaviour of the
gravity-Ising system, we use periodic boundary conditions by
identifying the top and bottom spatial slices of the cylindrical
histories contributing to the state sum \rf{2.1}.  Clearly this is not
going to affect the local properties of the model. A geometry
characterized by a toroidal triangulation $T$ of volume $N_T$ contains
$N_T$ time-like links, $N_T/2$ space-like links, $N_T/2$ vertices and
$3 N_T/2$ nearest-neighbour pairs.

The partition function of $n$ Ising models coupled to 2d Lorentzian
quantum gravity is given by
\beq{2.2}
G(\l,t,\b) = \sum_{T \in {\cal T}_t} e^{-\l N_T} Z_T^n(\b),
\eeq
where $T$ is now a triangulation with toroidal topology.  The
partition function for a single Ising model on the triangulation $T$
is denoted by $Z_T(\b)$, where the spins are located at the vertices
of $T$ and $\b$ is the inverse temperature of the Ising model.

On a fixed lattice there are no interactions among the $n$ Ising spin
copies if the partition function is simply taken as the $n$-fold
product of $Z(\b)$ for a single Ising model.  In the presence of
gravity, given by the definition \rf{2.2}, the situation is
different. Although the spin partition function $Z_T^n(\b)$ still
factorizes for any given $T$, this is no longer the case after the sum
over $T$ has been performed. The different spin copies are effectively
interacting via the triangulations (or in a continuum language: via
the geometry); the weight of each triangulation is a function of all
the $n$ Ising models.

It is straightforward to perform computer simulations of the combined
gravity-matter system given by \rf{2.2} (see \cite{aal} for details).
The only non-trivial aspect of the Monte Carlo simulation is the
updating of geometry, and for this the procedure used in \cite{aal}
can readily be generalized to the extended spin system of \rf{2.2}.
All results discussed in the following have been obtained at the
critical coupling $\b$ of the combined gravity-matter system \rf{2.2},
with $n\equ 8$, i.e. with central charge $c\equ 4$.  Our motivation
for choosing $c\equ 4$ comes from our experience with Euclidean
quantum gravity coupled to matter fields. In that case the phase
transition at $c=1$ is not very clearly visible in simulations. Only
for $c \geq 4$ can the changes in geometry be detected easily. We have
therefore chosen to work with 8 Ising spins in Lorentzian quantum
gravity, to have both $c$ sufficiently large to detect potential
effects on the geometry, but still small enough to make computer
simulations feasible within a limited amount of time.

\section{Numerical results}

We have performed our simulations on dynamically triangulated 
surfaces of torus topology with $N_T$ triangles (corresponding
to $N\equ N_T/2$ vertices) and $t$ time slices. 
For reasons that will become apparent in the following we have
used geometric configurations with different ratios of temporal 
length $t$ versus average spatial extent, satisfying 
$N=t^2/\t$ with $\t = 1,2,3$ and $4$. 
The choice $\t=1$, previously used in \cite{aal}, corresponds to 
a square lattice (with opposite sides identified),
while for $\t>1$ one obtains tori elongated in the $t$-direction.
The system sizes $N$ used in the simulations at various values
of $\t$ are listed in Table \ref{t:3:1}.
\TABULAR{|c|c|c|c|}{
\hline
$\t$=1 & $\t$=2 & $\t$=3 & $\t$=4 \\
\hline
 1024 & 1058 &  1200 & 1156 \\
 2025 & 2048 &  2352 & 2116 \\
 4096 & 4232 &  4800 & 4624 \\
 8100 & 8192 &  9408 & 8464 \\
16384 &      & 19200 &      \\
32400 &      & 36963 &      \\
\hline
}
{System sizes in terms of the number $N$ of vertices used in
simulations at $\t\equiv t^2/N\equ 1,2,3,4$.\label{t:3:1}} 
The geometry is updated using the move described in \cite{aal},
and for each geometry update (corresponding to approximately $N$ 
accepted moves) the Ising spins are updated with the
Swendsen--Wang algorithm. The focus of our attention is on the
multiple Ising model with $c\equ 4$, 
although for comparison
some data for $c\equ 0, 1/2$ will also be reported.

The first step of the numerical simulation consists in
determining the critical values $(\l_c,\b_c)$ of the
cosmological and the matter coupling constants. For the 
pure gravity model ($c\equ 0$), the cosmological constant
$\l_c\equ\ln 2$ is known exactly \cite{al}. For a single
Ising model ($c\equ 1/2$), we know from our previous simulations 
that $(\l_c,\b_c)\equ (0.742(5),0.2521(1))$ \cite{aal}, 
where the normalization for $\l_c$ is such that 
$\l_c=\ln 2$ at $\b=\infty$. For the case of eight Ising models,
using finite-size scaling as in \cite{aal}, for system sizes
$N\equ 1K$--$8K$ and $\t\equ 1,3$ we have obtained 
$(\l_c,\b_c)\equ (1.081(5),0.2480(4))$. As expected, this result is
insensitive to the value of $\t$.
Having established the critical values, we investigate 
finite-size scaling of the system at $(\l_c,\b_c)$. 
The statistics vary, for example, we performed
$1.88\times 10^6$ sweeps for the $N\equ 19200$ system and $0.6\times
10^6$ sweeps for the $N\equ 36963$ system. 
The data are binned for errors. 

We apply finite-size scaling to a variety of observables, 
in order to extract universal properties 
which characterize both the quantum geometry and the matter 
interacting with it.
The first of them involves a measurement of the distribution 
$SV(l)$ of spatial volumes (c.f. \cite{aal}), that is, the
lengths $l$ of slices at constant time $t$.
For sufficiently large lengths $l$ and space-time volumes $N$,
one expects a finite-size scaling relation of the form 
\beq{3:sv}
SV_N(l) = F_S(l/N_{T}^{1/\d_h}),
\eeq
for some function $F_{S}$. If such a relation holds, it defines
a relative dimensionality of space (characterized by the
average length $\la l \ra$) and (proper) time since from
\beq{3:tl}
N_T \sim t \cdot \la l\ra \; \Longrightarrow \; 
t \sim N_T^{1-1/\d_h},\quad \la l\ra \sim N_T^{1/\d_h}.
\eeq
By relating the geodesic distance $t$ in time direction to
the total volume, we can define a global or cosmological Hausdorff 
dimension $d_{H}$ of space-time through
\beq{3:td}
t \sim N_T^{1/d_H}\;\Longrightarrow \; d_H =  \frac{\d_h}{\d_h-1}.
\eeq
This definition is motivated by a similar notion in Euclidean 
quantum gravity. In that case there is no distinction
between spatial and time directions, and one can extract the
global Hausdorff dimension by measuring the volumes of spherical
shells at geodesic distance $r$ (the analogue of the geodesic
time $t$ above) from a given point. (Note that a ``shell'' need not
be a connected curve.) 
In a discretized context this amounts to counting the number 
$ n_N(r)$ of vertices at geodesic (link) distance $r$.
For this quantity one expects a 
scaling behaviour \cite{syr,ajw} of the type
\beq{3:nv}
 n_N(r) =N_T^{1-1/d_H}F_1(x),\qquad x = \frac{r}{N_T^{1/d_H}},
\eeq
which has been verified for the case of 2d Euclidean quantum 
gravity.
Eq.\ \rf{3:nv} is a typical example of a finite-size scaling relation. 
It tells us how a radial or proper time coordinate has to scale 
with space-time volume in order to obtain a non-trivial continuum limit 
($N \to \infty$, $r \to \infty$). In this sense $d_{H}$ describes
long-range properties of the system, which is our rationale for
calling it the {\it cosmological} Hausdorff dimension. 
It does not necessarily tell us about the short-distance behaviour 
of space-time, for example, how
the volume of a spherical shell behaves at 
small radius $r \ll N_T^{1/d_H}$ (but still with $r$ much larger 
than the lattice spacing, to avoid lattice artifacts). 
At such distances one expects the shell volume to grow with a
power law
\beq{3:dh}
n_N(r) \sim r^{d_h-1}, 
\eeq
where $d_h$ is now a ``short-distance'' fractal dimension \cite{www}. 
There is no {\it a priori} reason for $d_{h}$ to coincide with
the cosmological Hausdorff dimension. However, in models of
simplicial Euclidean quantum gravity we have always observed 
$d_h\equ d_H$, such that \rf{3:nv} was valid for all $r$ 
(much larger than the lattice spacing). 
This points to a unique fractal structure of space-time, with 
$F_1(x)\equ x^{d_h-1}$ for $x \ll 1$. Nevertheless, there 
also exist related models with $d_h \neq d_H$ \cite{check}.
We will see below that Lorentzian gravity coupled to sufficiently
much matter provides another example of this kind.

For illustrative purposes we have generated 
3D visualizations of the two-dimen\-sio\-nal
dynamically triangulated geometries produced during the simulations. 
For {\it Lorentzian} geometries this can easily be done: 
as a consequence of the
causality requirement each 2d history consists of an
ordered sequence of 1d spatial slices of constant time.
Each such slice is embedded isometrically in three-dimensional 
flat space and then the vertices of
neighbouring slices are connected. Different colours 
indicate clusters of spin-up and spin-down states. For the
case of multiple Ising models, one of them is chosen arbitrarily
to determine the surface colouring. 
We have cut open the toroidal geometries along one of their 
spatial slices,
so that in the pictures they appear as cylinders (with
top and bottom slices to be identified).
The visualizations are well suited for comparing 
the qualitative behaviour of the geometric and spin degrees
of freedom as well as their interaction, for different
values of the conformal charge $c$. 
Animations
of some of the simulations can be found at \cite{aalm}.

\subsection{Lorentzian quantum gravity with $c\leq 1/2$}

To put our current results into context, let us recall the situation 
for pure Lorentzian gravity ($c \equ 0$) and 
for Lorentzian gravity coupled to one critical Ising model ($c \equ 1/2$).
In that case, independent measurements of $SV_N(l)$ and $n_N(t)$
both yield $\d_h \equ d_H \equ 2$, corroborating the existence
of a universal fractal dimension $d \equ 2$, which moreover
coincides with the na\"\i vely expected continuum value.
In addition, we have found the Onsager exponents for the case of
a single Ising model coupled to Lorentzian gravity.
The fact that both the fractal dimensions and the critical matter
exponents retain their ``canonical'' values is
in sharp contrast with the situation in 2d dynamically triangulated
{\it Euclidean} quantum gravity. 

For later comparison with the case of eight
Ising models coupled to Lorentzian gravity, 
we show in Fig.~\ref{f:3:5a} two typical configurations  
of the pure-gravity system for $N\equ 8100$, $\tau\equ 1$, and for
$N\equ 9408$, $\tau \equ 3$, generated during the Monte Carlo simulations. 
Apart from an overall rescaling, the $\t\equ 1$ geometry looks 
qualitatively similar to its $\t\equ 3$ counterpart. 
This observation can be made into a quantitative statement by showing 
that the distribution $SV_N(l)$ is independent of $\t$, as indeed we
have done. 
\FIGURE{
\def\xsize{2.4in}
\def\ysize{2.4in}
\epsfxsize=\xsize \epsfysize=\ysize 
\epsfbox{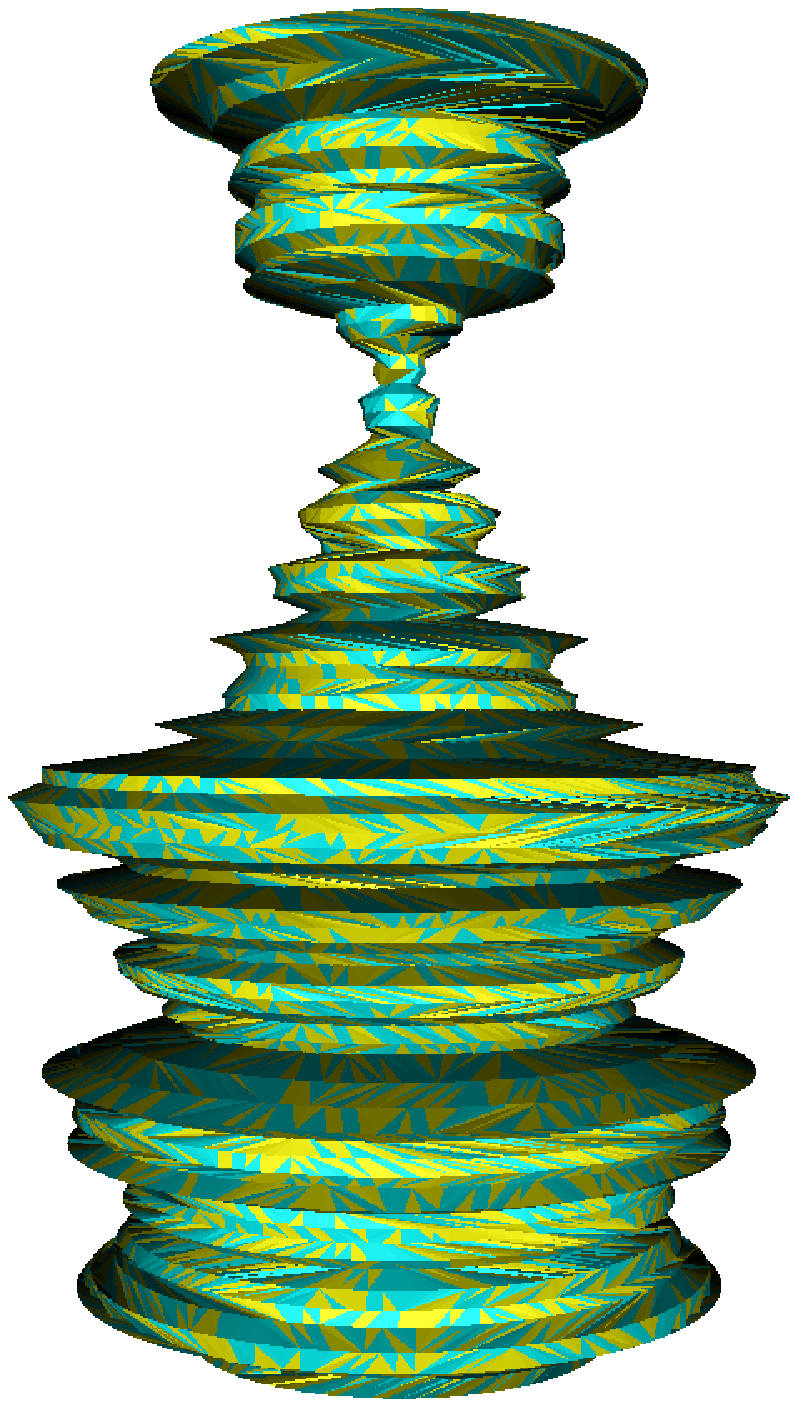}
\epsfxsize=\xsize \epsfysize=\ysize 
\epsfbox{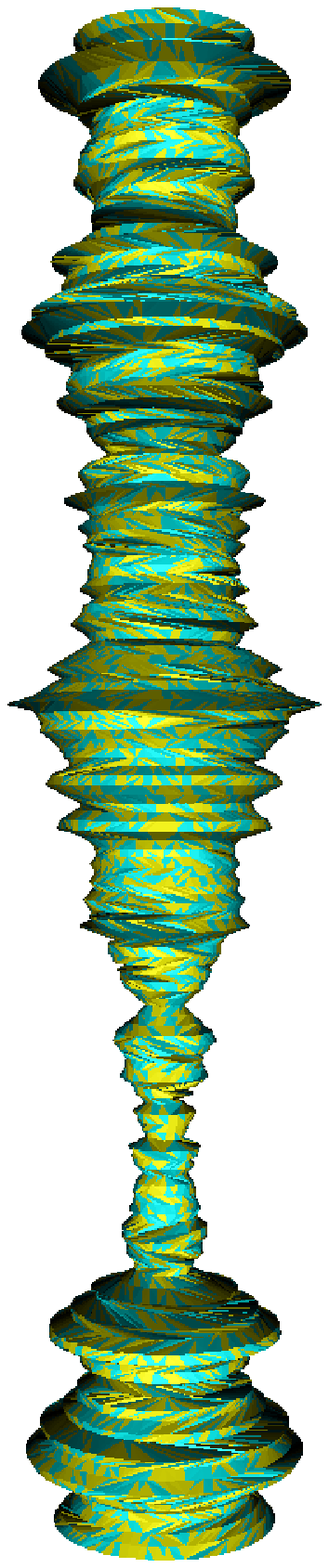}
\caption{Typical configurations for $c\equ 0$, with volumes 
$N\equ 8100$ ($\t\equ 1$) and $N\equ 9408$ ($\t\equ 3$). 
The spin configurations are those of a $\b\equ 0$ system.}
\label{f:3:5a}
}%
From the point of view of the space-time geometry, the situation is 
similar for $c\equ 1/2$ coupled to Lorentzian 
gravity. We illustrate this by two typical configurations, 
depicted in Fig.\ \ref{f:3:7}. Also in this case we have
checked that the distribution $SV_N(l)$ is independent of the
relative temporal extension $\t$ of space-time.  
\FIGURE{
\def\xsize{2.4in}
\def\ysize{2.4in}
\epsfxsize=\xsize \epsfysize=\ysize 
\epsfbox{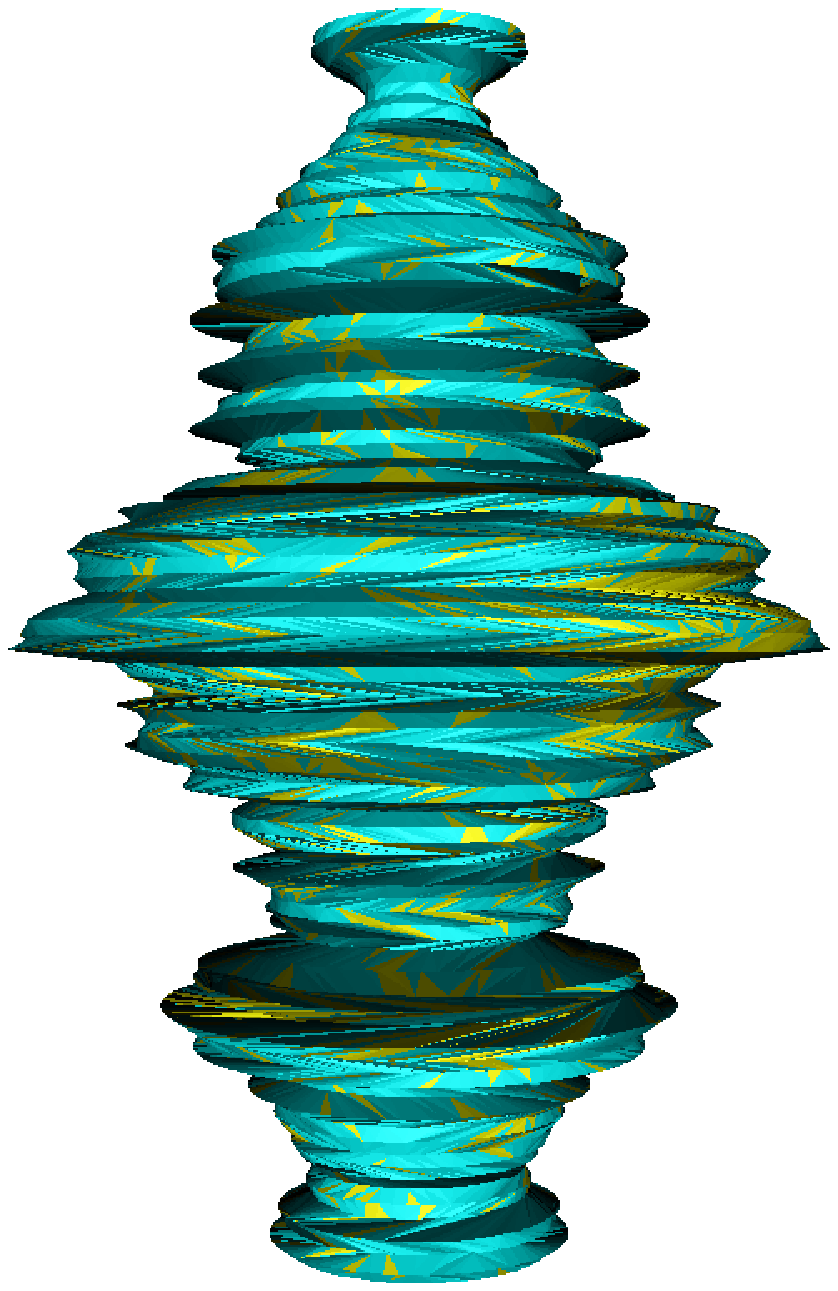}
\epsfxsize=\xsize \epsfysize=\ysize 
\epsfbox{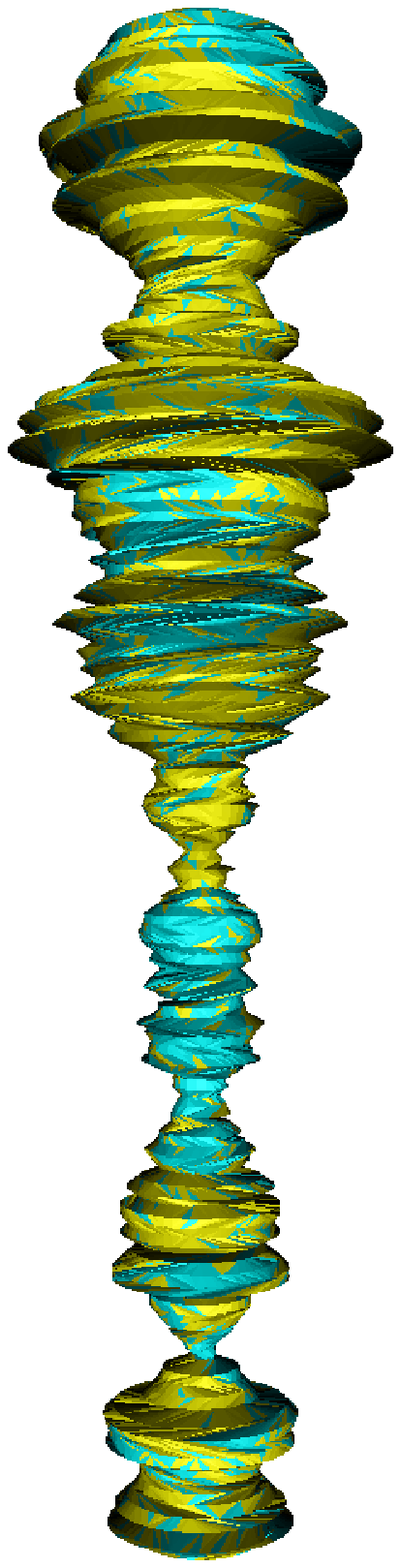}
\caption{Typical configurations for $c\equ 1/2$, with volumes 
$N\equ 8100$ ($\t\equ 1$) and $N\equ 9408$ ($\t\equ 3$).}
\label{f:3:7}
}%
 
\subsection{Properties of the quantum geometry for $c=4$}

\subsubsection{The length distribution $SV_N(l)$ and 
the dimension of proper time}

In the same manner as discussed above, we can
extract some large-scale characteristics of the quantum
geometry of the $c \equ 4$ system coupled to Lorentzian gravity
by studying the scaling properties of the distribution
$SV_N(l)$ of one-dimensional spatial slices of volume $l$. 
It turns out that for $c\equ 4$ one has to simulate systems
with $\t\ge 3$ to observe a clear scaling behaviour. 
As illustrated by Fig.~\ref{f:3:4},%
\FIGURE{
\def\xsize{2.9in}
\def\ysize{1.93in}
\epsfxsize=\xsize \epsfysize=\ysize 
\epsfbox{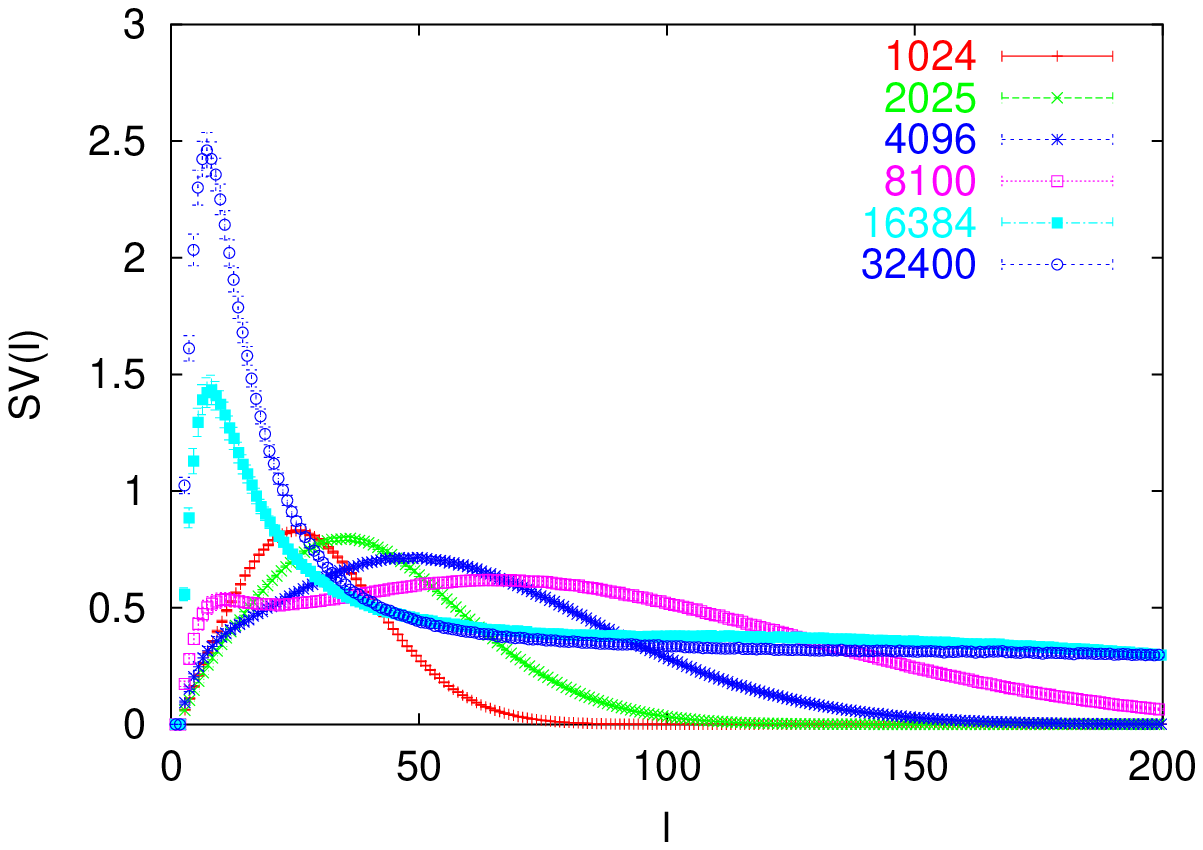}
\epsfxsize=\xsize \epsfysize=\ysize 
\epsfbox{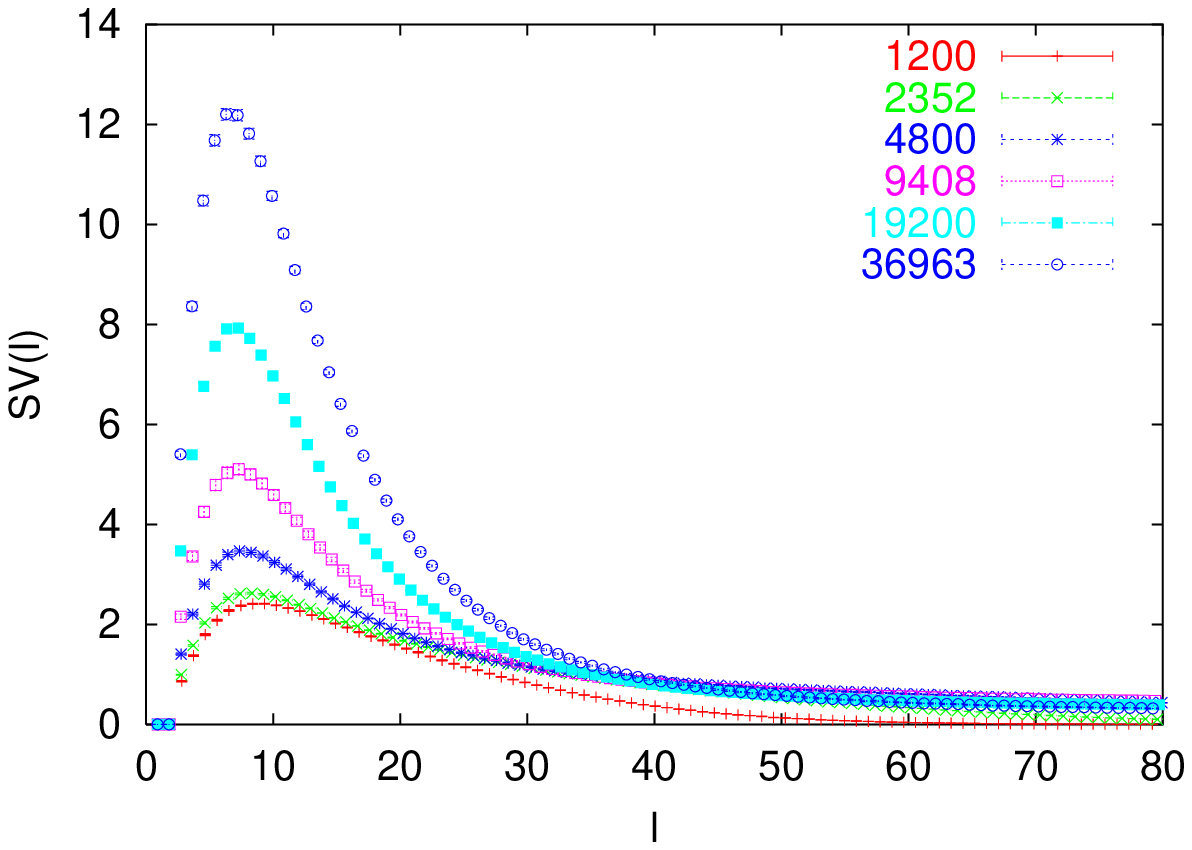}
\caption{The appearance of spatial slices of very short length 
for $c\equ4$, and for $\t\equ 1$ (left) and $\t\equ 3$ (right).} 
\label{f:3:4}
}%
the system exhibits a tendency for developing a large number
of very short spatial slices, 
with length of the order of the cut-off. 
The length distribution $SV_N(l)$ has a peak at small $l$, whose
height increases with $N$, but whose position has a very weak 
dependence on the system size. 
Since the volume is kept fixed, there are strong finite-size
effects which artificially prohibit the system from forming such a
peak whenever $N$ and $\t$ are simultaneously small. This is
obvious from the data taken for the $\t\equ 1$ system (Fig.~\ref{f:3:4}).
In that case the peak appears clearly only for a system with
more than $8100$ vertices. 
Such finite-size effects are absent for the $\t\equ 3$ system. 

These properties are well illustrated by
Figs.\ \ref{f:3:5} and \ref{f:3:8}, which 
show some typical geometries at $c\equ 4$. 
They should be compared to our previous
Figs.\ \ref{f:3:5a} and \ref{f:3:7} for $c\leq 1/2$.
\FIGURE{
\def\xsize{2.4in}
\def\ysize{2.4in}
\epsfxsize=\xsize \epsfysize=\ysize 
\epsfbox{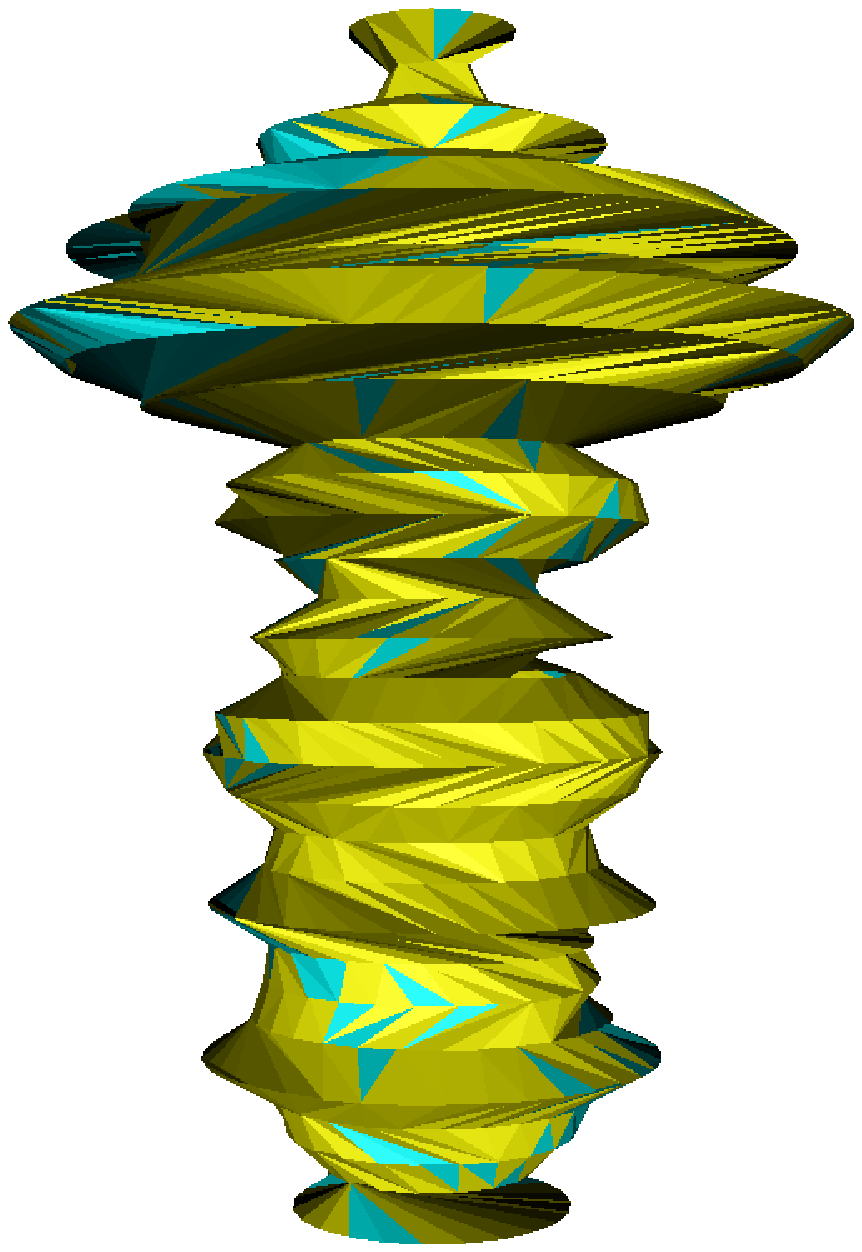}
\epsfxsize=\xsize \epsfysize=\ysize 
\epsfbox{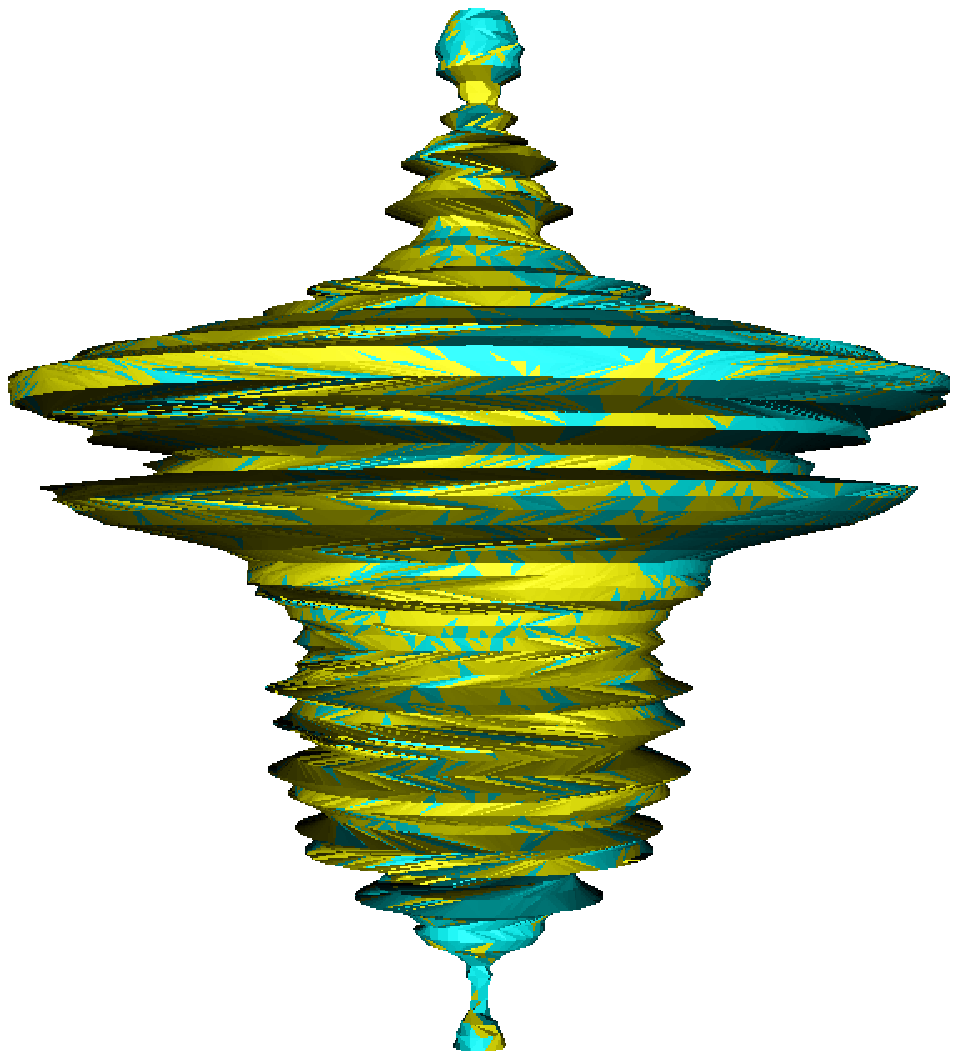}\\
\epsfxsize=\xsize \epsfysize=\ysize 
\epsfbox{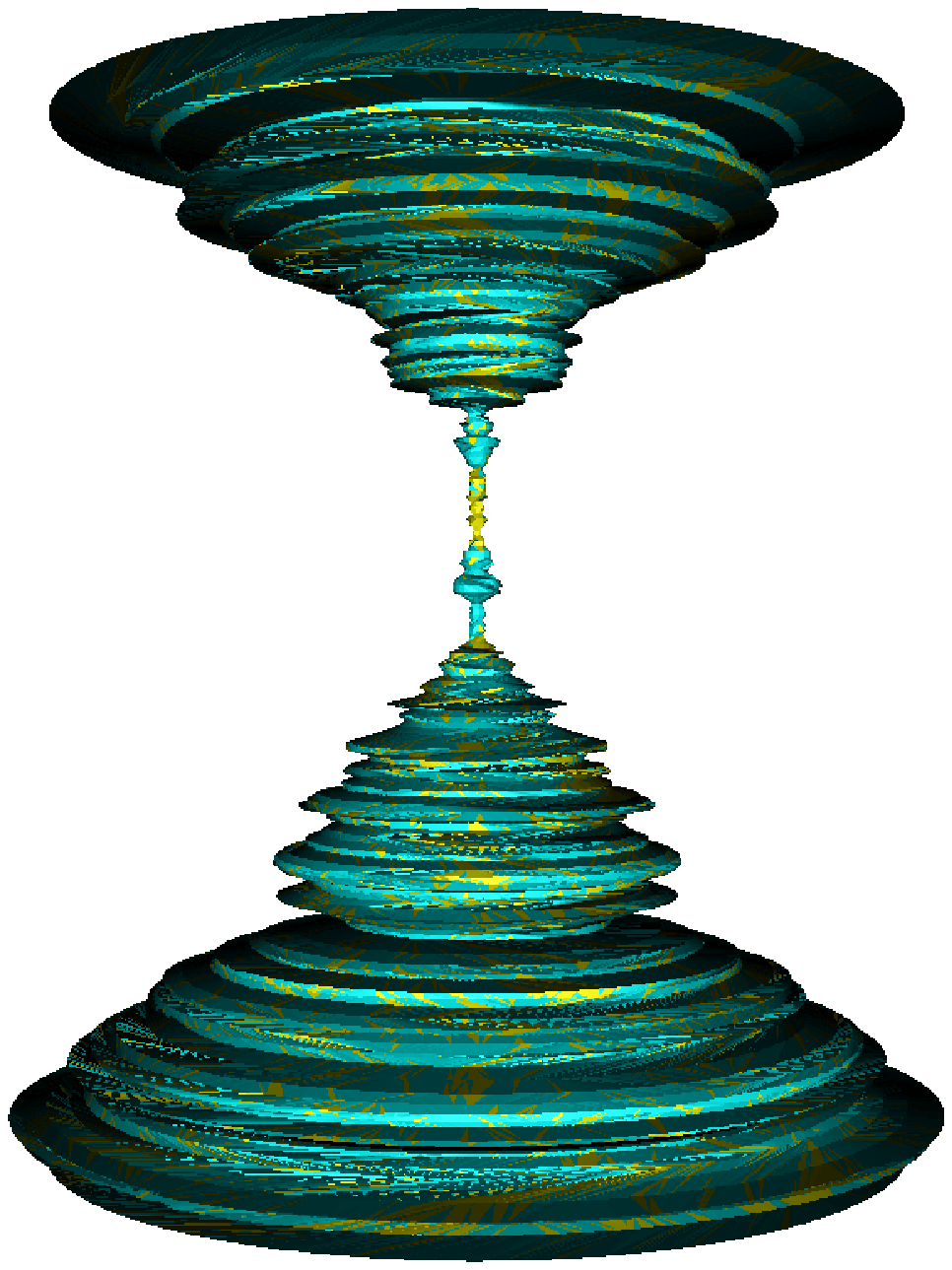}
\caption{Typical configurations for $c\equ 4$, $\t\equ 1$ and 
volumes $N\equ 1024,8100$ and $32400$. }
\label{f:3:5}
}%
Fig.\ \ref{f:3:5} contains space-time configurations at 
$\t \equ 1$, for three different volumes $N$. Their tendency to 
separate into two distinct regions increases with $N$ (remember that 
the time direction has been chosen periodic).
This is a typical finite-$N$ behaviour associated with a phase
transition, in this case, of the geometry.
Likewise, for increasing $\t$ (and constant volume) 
it becomes easier to form long and thin ``necks'',
along which the spatial volumes $l$ stay close to the
cut-off size (see Fig.\ \ref{f:3:8}).
Note in particular the space-time 
history with the largest volume ($N\equ 36963$), 
where the separation of space-time into two different phases 
is very pronounced, underscoring at the same time the effect
of increasing $N$.
This figure also illustrates the fact that in the limit as
$N \to \infty$, the neck region will carry a vanishing
space-time volume.

It happens only rarely that the extended region shows a tendency 
to break up into smaller parts. Generally speaking, the 
fluctuations in its shape constitute the slowest modes of
the simulation. Occasionally we observe a (much)
smaller extended region splitting off from the main one. 
However, our statistics was insufficient to establish 
whether for large $\t$ there is an underlying pattern governing 
the size and frequency of these events. 
For our present purposes, this effect can be
safely ignored, since the number and size of such secondary
space-time regions was small.

\FIGURE{
\def\xsize{2.4in}
\def\ysize{2.4in}
\epsfxsize=\xsize \epsfysize=\ysize 
\epsfbox{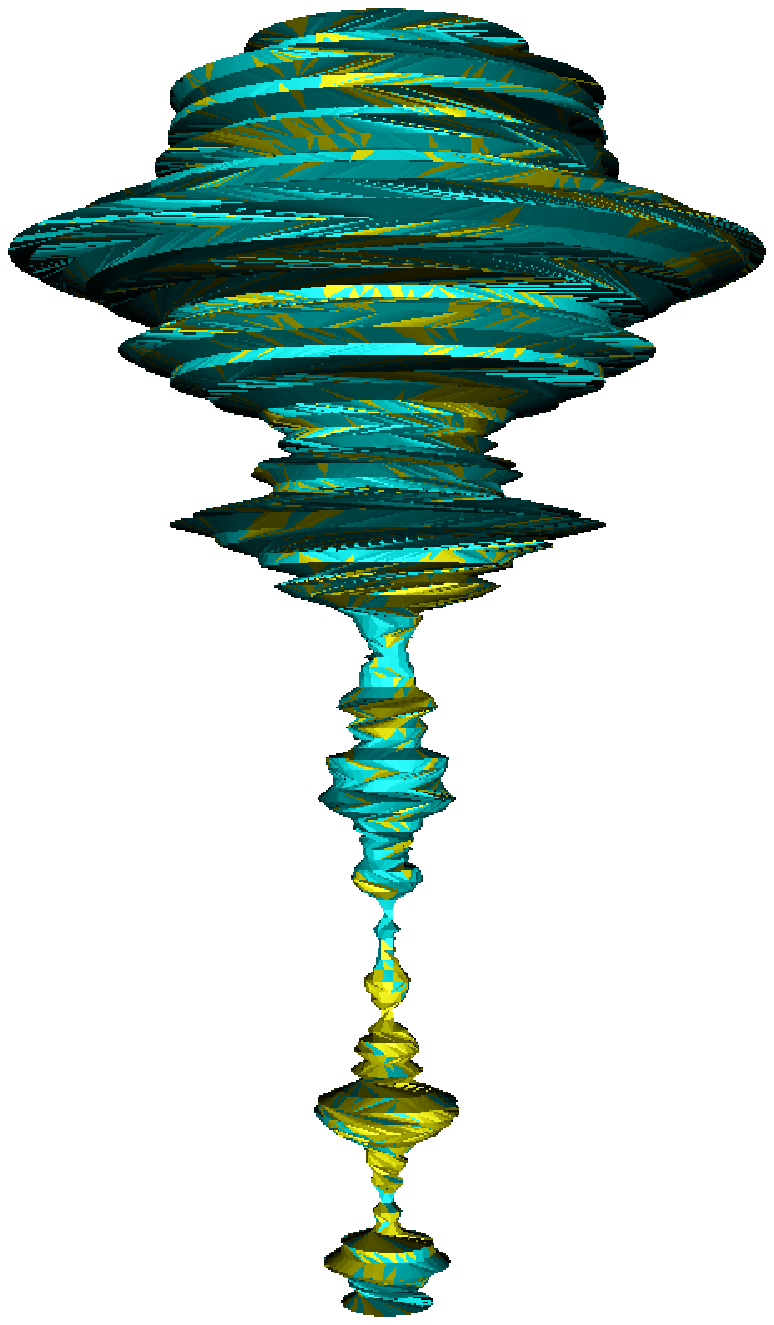}
\epsfxsize=\xsize \epsfysize=\ysize 
\epsfbox{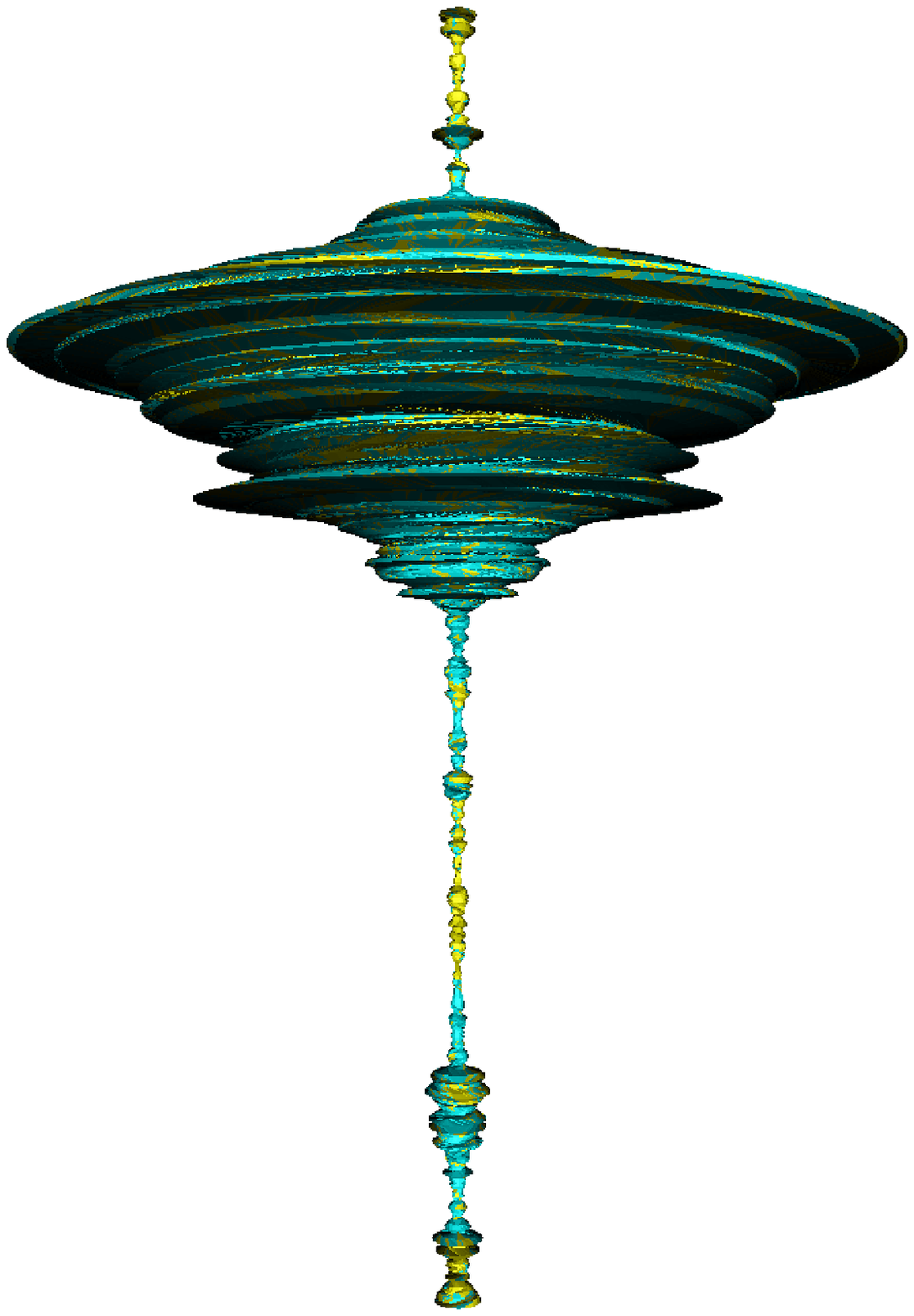}\\
\epsfxsize=\xsize \epsfysize=\ysize 
\epsfbox{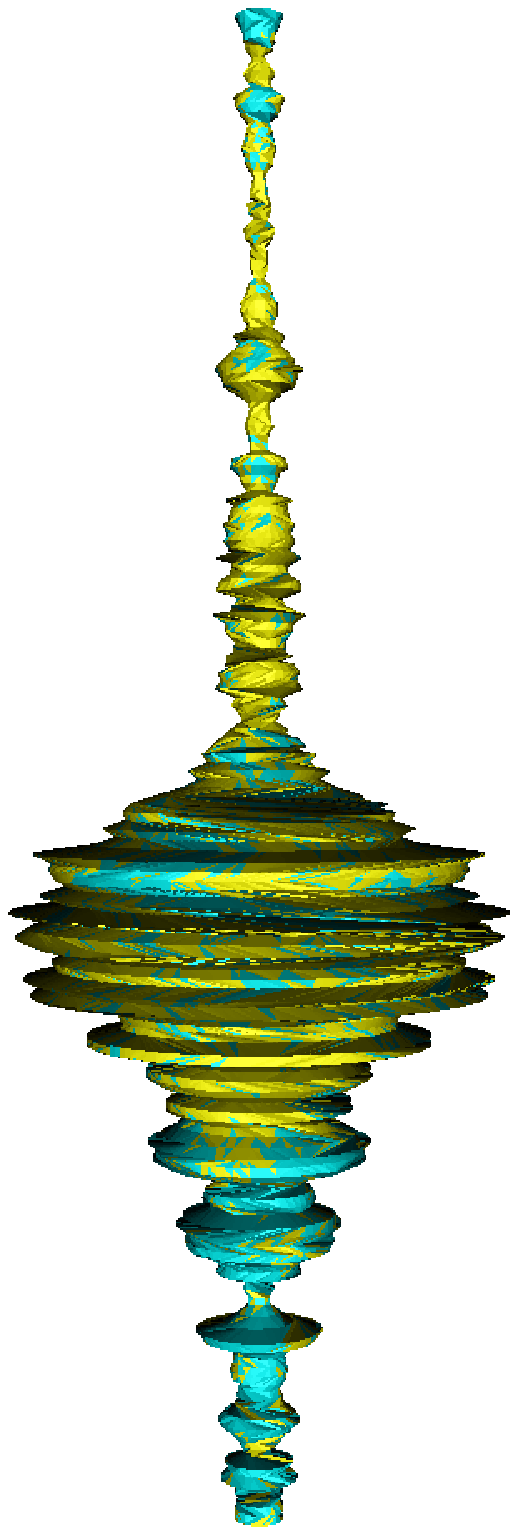}
\caption{Typical configurations for $c\equ 4$, with volumes 
$N\equ 8192$ ($\t\equ 2$), 
$N\equ 36963$ ($\t\equ 3$) and $N\equ 8464$ ($\t\equ 4$).}
\label{f:3:8}
}%

Let us now quantify the scenario just outlined 
by a study of the scaling properties of $SV_N(l)$.
As expected, the length distributions $SV_N(l)$ show no sign of 
scaling for small $l$. For large $l$, however, a scaling relation
of type \rf{3:sv} is well satisfied, as illustrated by the plots in 
Fig.~\ref{f:3:9}.%
\TABULAR{|c|c|c|c|}{
\hline
$\t$=1  & $\t$=2  & $\t$=3   & $\t$=4  \\
\hline
1.70(3) & 1.65(5) &  1.54(3) & 1.50(3) \\
\hline
}
{The optimal values of the exponent $\d_h$ for best scaling of $SV(l)$
fitting \protect\rf{3:sv}.\label{t:3:2}} 
\FIGURE{
\def\xsize{2.9in}
\def\ysize{1.93in}
\epsfxsize=\xsize \epsfysize=\ysize 
\epsfbox{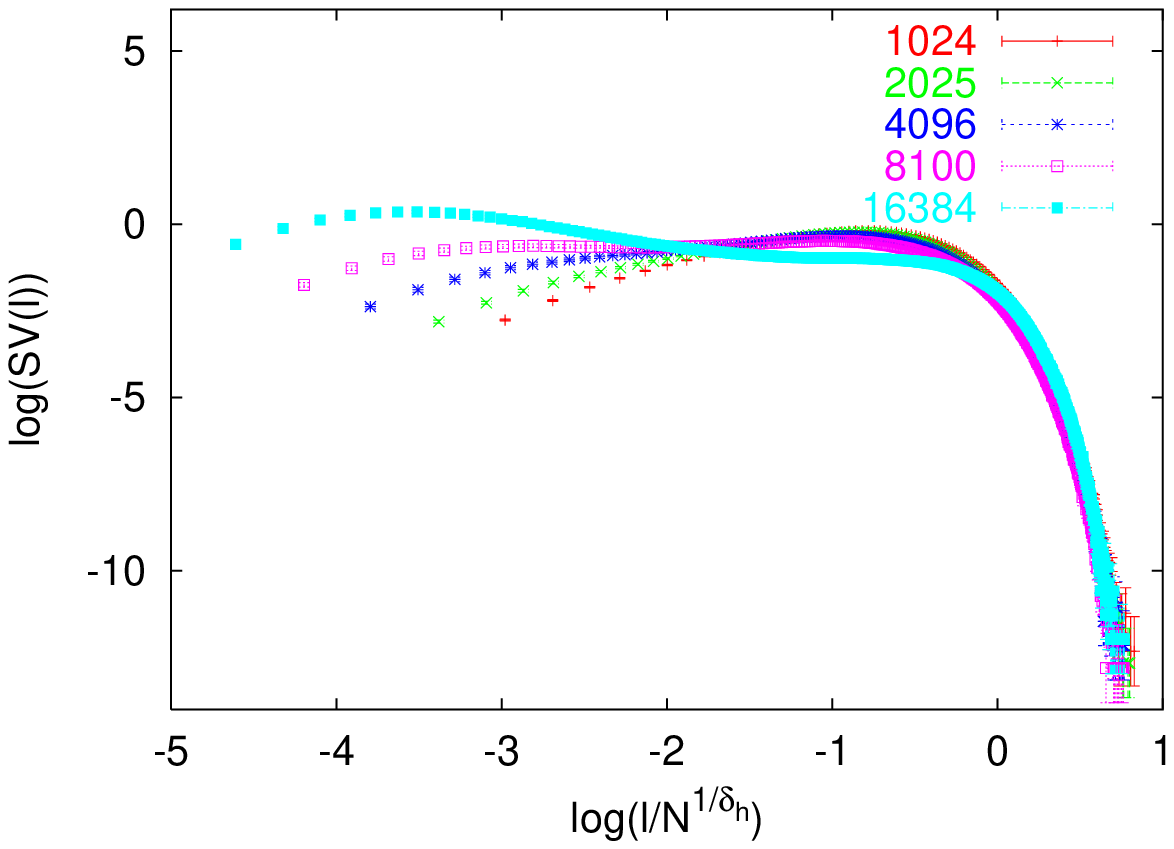}
\epsfxsize=\xsize \epsfysize=\ysize 
\epsfbox{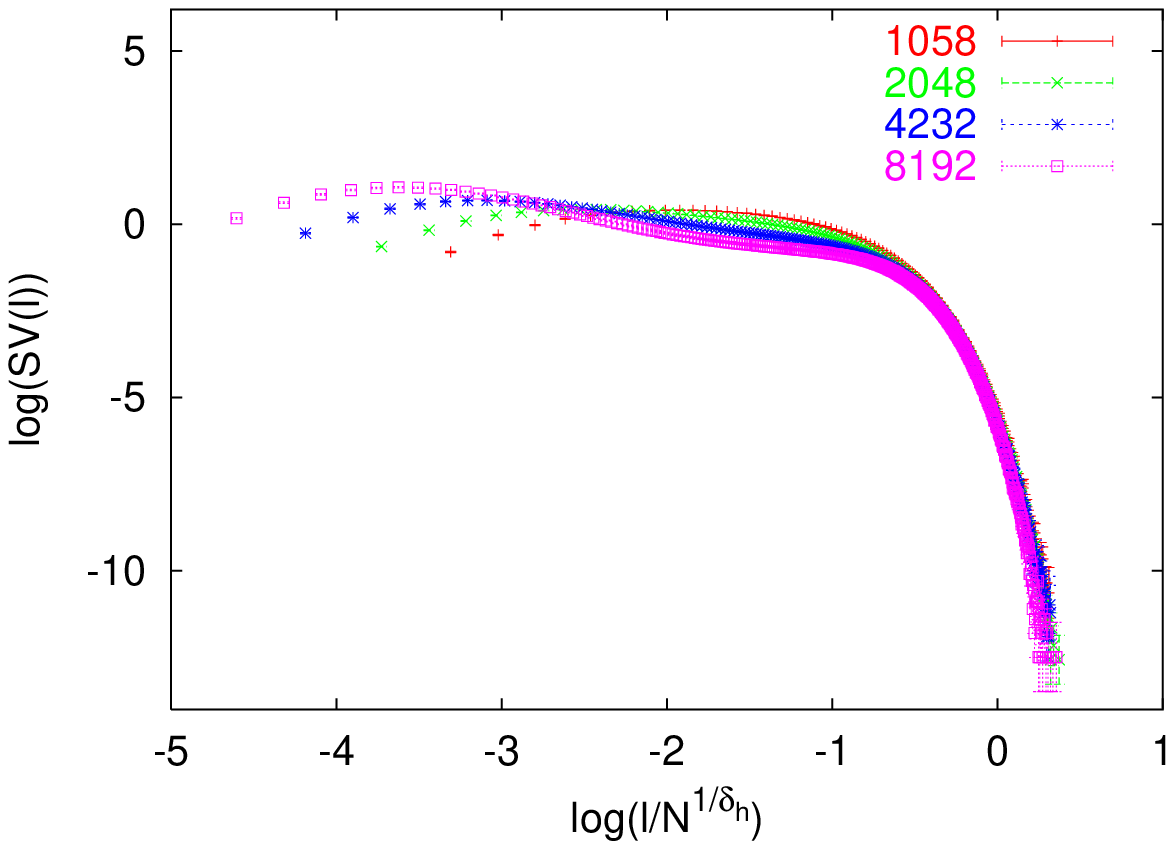}\\%
\epsfxsize=\xsize \epsfysize=\ysize 
\epsfbox{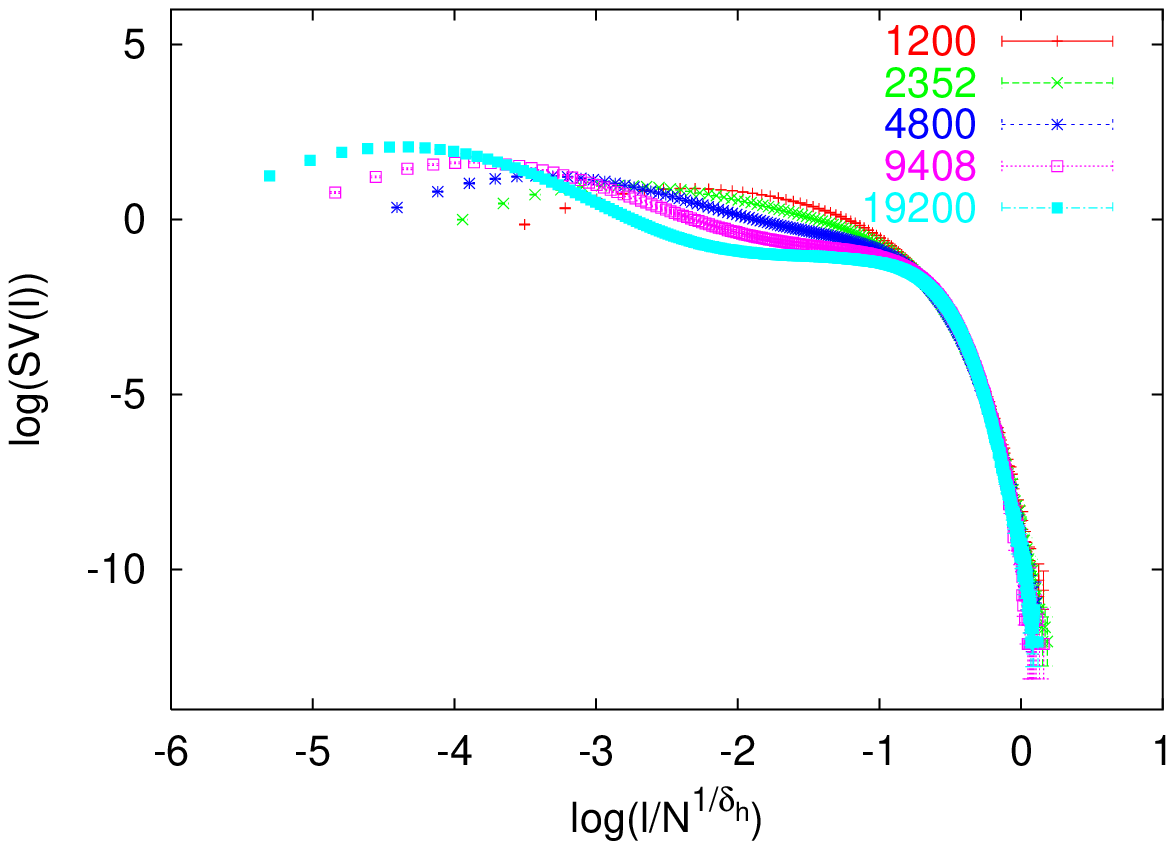}
\epsfxsize=\xsize \epsfysize=\ysize 
\epsfbox{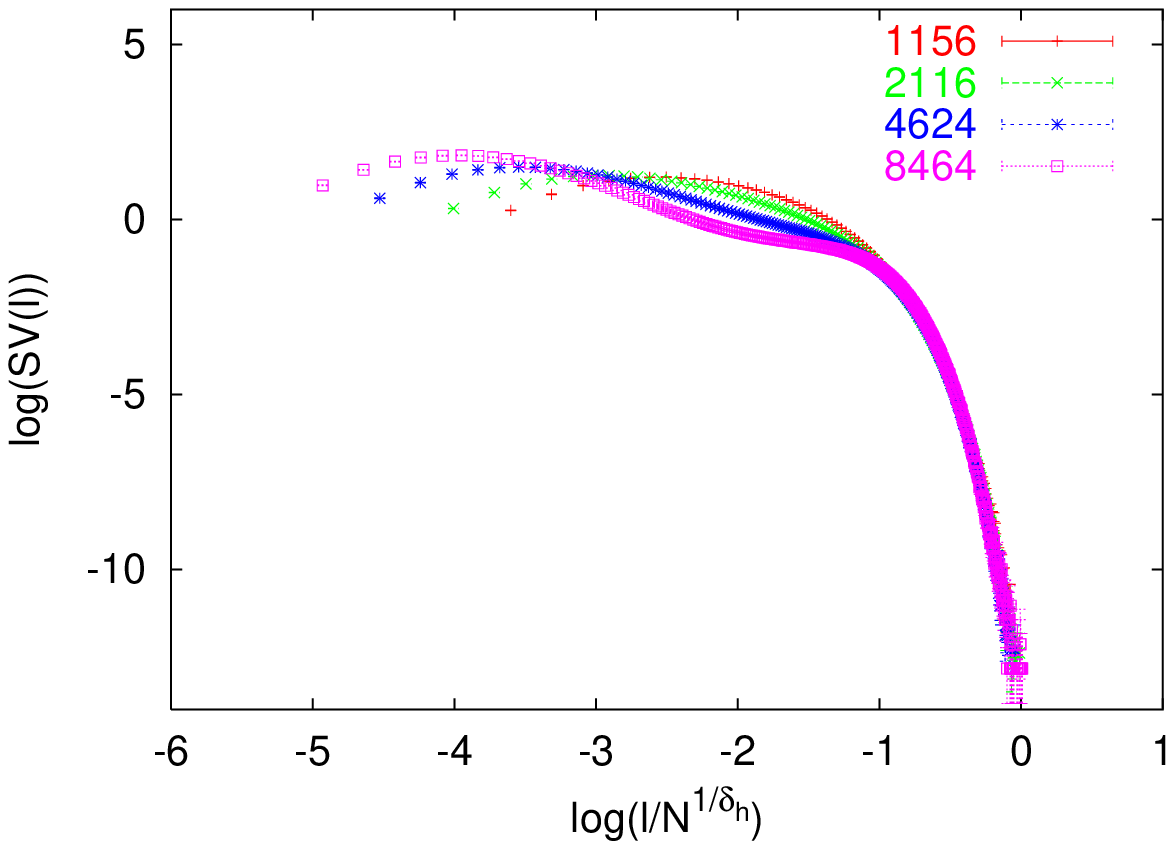}
\caption{The distributions $SV(l)$ for the $c\equ 4$, $\t\equ 1,2,3$ and $4$
systems.}
\label{f:3:9}
}%
The optimal values for $\d_h$ are contained in Table~\ref{t:3:1}. 
There is a clear tendency for $\d_h\rightarrow 3/2$ as $\t$ becomes
large. From Fig.\ \ref{f:3:9} we can read off at
which value of the parameter $x=l/N_T^{\delta_h}$ the scaling sets in. 
This happens for
$x \geq c$, where $c \approx 0.5$, or (setting $\d_h \equ 3/2$) 
for lengths
\beq{3:sca}
l \geq c\, N_T^{2/3}.
\eeq
As mentioned above, the neck region does not contribute significantly 
to the volume for large $N$. 
We have measured that the volume $V_{ext}$ of the
extended phase (now defined as the scaling region of $SV(l)$) 
is asymptotically proportional to the total volume $V$($\equiv N_T$) 
of the surface. 
Let $t_{ext}$ denote the temporal extension of this extended region
and $l_{ext}$ the typical length of a spatial slice in that region
(such that $t_{ext}\cdot l_{ext}=V_{ext}\propto N$). 
If we assume for the sake of definiteness that indeed $\d_h\equ 3/2$, 
it follows from \rf{3:td} and \rf{3:tl} that  
\bea
{\rm dim}\, V_{ext} & = & \frac{3}{2}\, {\rm dim}\, l_{ext}.  \label{3:dlv} 
\eea
From this we immediately deduce the relations
\bea
   {\rm dim}\, l_{ext} & = & 2 \, {\rm dim}\, t_{ext},    \label{3:dlt} \\
   {\rm dim}\, V & = & 3 \, {\rm dim}\, t_{ext}   \label{3:dtv}, \, 
\eea
and that the cosmological Hausdorff dimension
is given by $d_H\equ 3$.

Our main conclusion is that the coupling of 8 Ising models to 
Lorentzian gravity produces a phase transition in which some
universal properties of the geometry are changed. 
At large distances, proper time and 
spatial length develop anomalous dimensions relative 
to each other and to the space-time volume, as expressed by 
eqs.\ \rf{3:dlt} and \rf{3:dtv}.   

\subsubsection{The shell volume $n_N(r)$ and the short-distance 
dimension $d_h$}

Next we discuss the measurement of the one-dimensional volumes 
$n_N(r)$ of spherical shells at distance $r$. 
It turns out that for $c\equ 4$ Lorentzian gravity plus matter,
these functions do not exhibit the universal scaling
properties found elsewhere in models of two-dimensional gravity
\cite{syr,ajw,check}. 
(That a universal behaviour at all length scales is unlikely is
already illustrated by the separation of typical configurations 
into a thin and an
extended region apparent in Fig.\ \ref{f:3:8}.) 
As discussed at the beginning of this section, this is no reason
for concern; it simply reflects the fact that the underlying
quantum geometry is more complex. 
We will identify several well-defined scaling regions 
and encounter the more general situation where the short-distance 
and the cosmological Hausdorff dimensions are different.

As can be seen in Fig.\ \ref{f:3:10}, 
the short-distance behaviour of $n_N(r)$ is independent
of $N$ and can be fitted nicely to $n_N(r) \sim r^{d_h-1}$, 
with $d_h \approx 2$ (which coincides with the value found
for $c\equ 1/2$ and also happens to be the ``canonical"
dimension expected from classical considerations). 
The best fit gives $d_h=2.1(2)$.
\FIGURE{
\def\xsize{4.0in}
\def\ysize{2.67in}
\epsfxsize=\xsize \epsfysize=\ysize 
\epsfbox{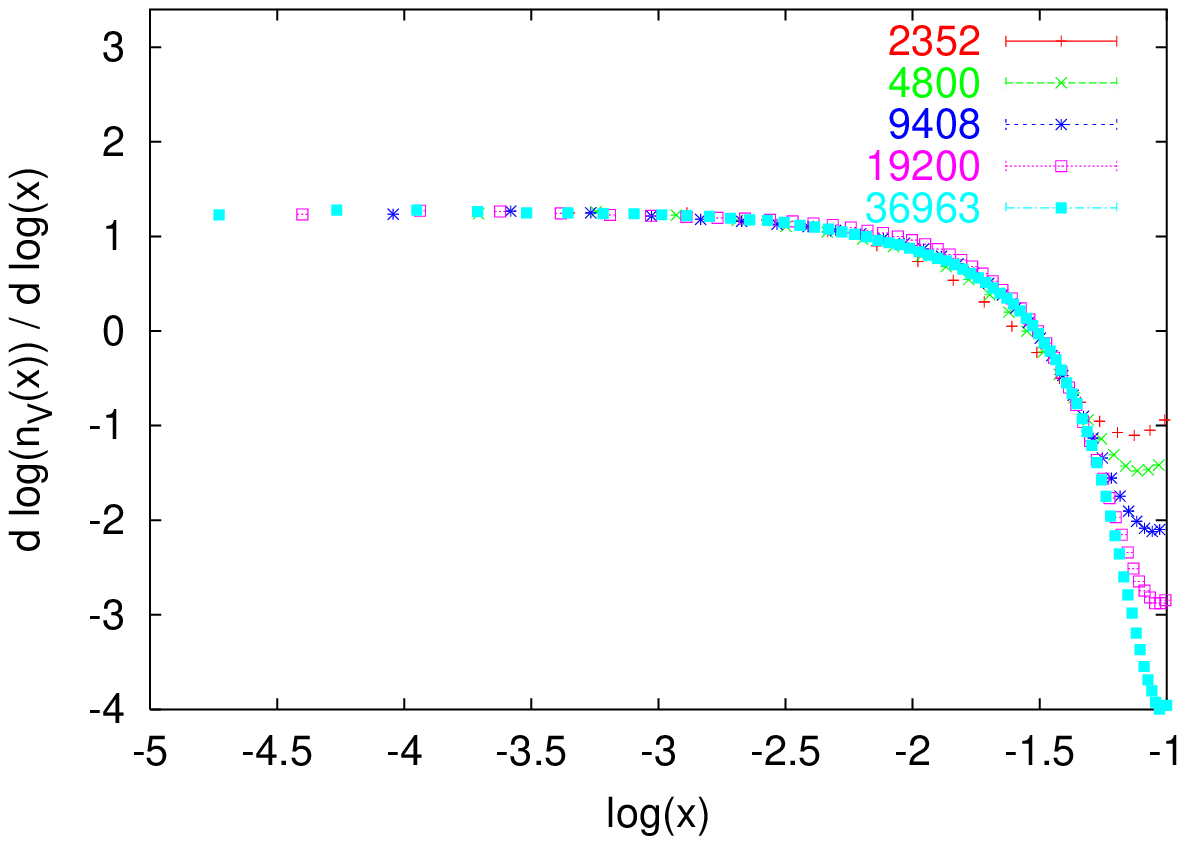}
\caption{Small-distance scaling of the correlation function
$n_N(r)$ for the $c\equ 4$ system.} 
\label{f:3:10}
}%

Going out to length scales of the order $r \sim N^{1/3}$, we see a 
different scaling behaviour. Here \rf{3:nv} is valid with a 
cosmological Hausdorff dimension $d_H \equ 3$, in accordance with 
the value extracted from the 
measurements of the length distribution $SV_N(l)$. 
Finite-size scaling in this region, computed from the scaling of 
the peaks of $n_N(r)$ (Fig.~\ref{f:3:11}), yields $d_H=3.07(9)$. 
\FIGURE{
\def\xsize{4.0in}
\def\ysize{2.67in}
\epsfxsize=\xsize \epsfysize=\ysize 
\epsfbox{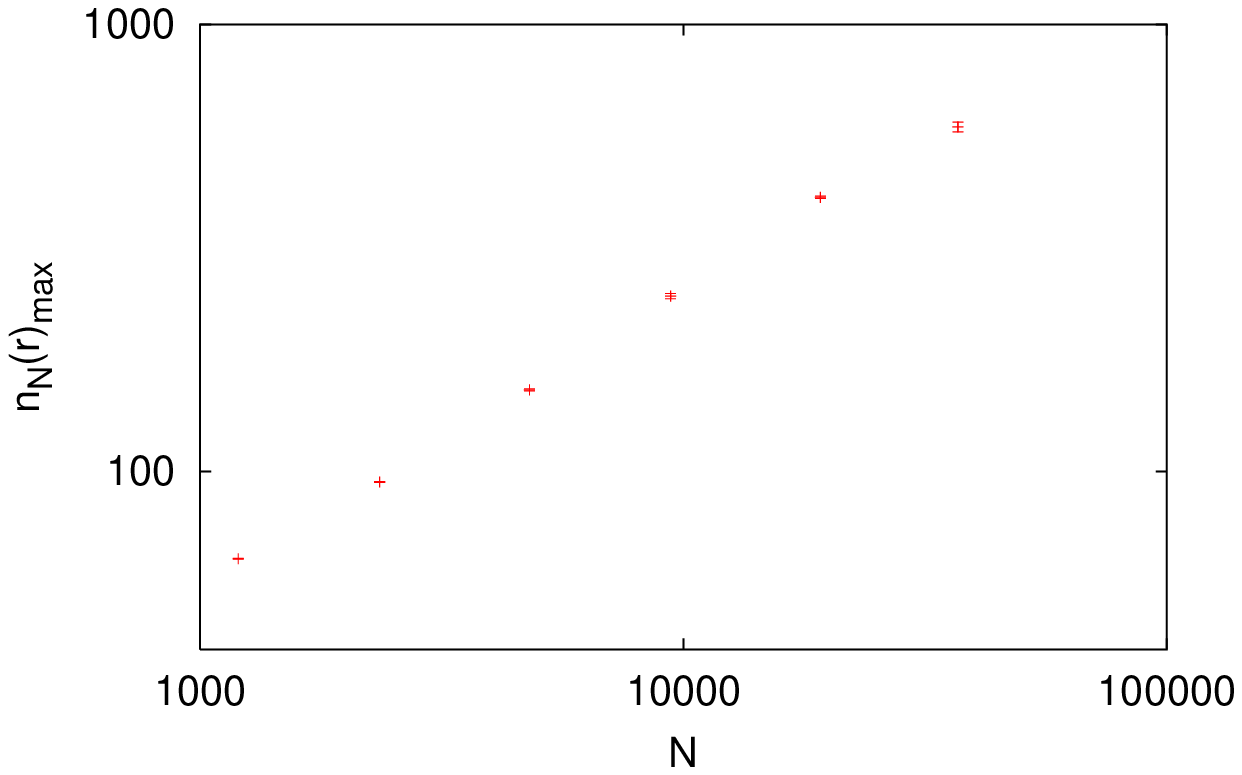}
\caption{Scaling of the peaks of the correlation function $n_N(r)$ 
for the $c\equ 4$ system.}
\label{f:3:11}
} 

Finally, at very large $r$ near the tail of the 
distribution, we found that the value of $n_N(r)$ is 
almost independent of $N$, indicating a dominance of
configurations with $d_H\equ 1$. 
Recalling the typical shape of configurations at
$c\equ 4$ (Fig.\ \ref{f:3:8}), this suggests the following 
interpretation.
Each measurement of $n_N(r)$ involves the choice of a 
reference point, from which the geodesic distances $r$
are measured. Since almost no space-time volume
is contained in the thin necks, the randomly chosen reference 
point will typically be located somewhere in the extended region.
However, moving outwards from such a bulk point in spherical
shells will for large $r$ eventually bring us back to
the neck region, which in 
the large-$N$ limit has a length proportional to $\sqrt{N}$
(simply because $N\propto t^2$). 
Once the spherical shells have reached the neck region, 
the volume function $n_N(r)$ will just measure a one-dimensional 
structure.

\subsection{Matter behaviour in the extended phase}

We have seen above how a Lorentzian geometry separates 
into two distinct regions under the coupling to 8 Ising models.
Since the thin, stalk-like region is effectively one-dimensional,
a non-trivial matter behaviour can be expected only in the
remaining, spatially extended space-time region.
The computation of the matter exponents in this phase 
is subtle and requires some care. 
At the critical matter coupling $\b_c$ we have measured the same 
set of observables as in our previous simulations \cite{aal}.
Together with their expected finite-size scaling behaviour they are
\beq{chi}
\chi = N (\la m^2\ra -\la |m| \ra^2) \sim N^{\g/\n d_H}~~~~~~~~~~~~
{\rm (susceptibility)}
\eeq
\beq{lnm}
D_{\ln |m|}= N \Big( \la e \ra - \frac{\la e |m|\ra}{\la |m| \ra} \Big)
\sim N^{1/\n d_H}~~~~~~(D_{\ln |m|} \equiv \frac{d \ln |m|}{d \b})
\eeq
\beq{lnm2}
D_{\ln m^2} = N \Big(\la e \ra - \frac{\la e m^2\ra}{\la m^2 \ra} \Big)
\sim N^{1/\n d_H} ~~~~~(D_{\ln m^2} \equiv \frac{d \ln m^2}{d \b}), 
\eeq
where $\g$ and $\n$ are the critical exponents of the susceptibility
and of the divergent spin-spin correlation length. 

Initially we checked that for the case of a single Ising model,
extending the geometries in the temporal direction (i.e. taking
$\t>1$) does not affect the Onsager exponents found in \cite{aal}. 
The results are tabulated in Table~\ref{t:3:3}, and do not differ
significantly from our previous results. 
\TABULAR{|c|c|c|}{
\hline
Observable    & Exponent            &        \\
\hline
$\chi$        & $\frac{\g}{\n d_H}$ &0.84(1) \\
$D_{\ln |m|}$ & $\frac{ 1}{\n d_H}$ &0.552(5)\\
$D_{\ln m^2}$ & $\frac{ 1}{\n d_H}$ &0.550(4)\\ 
\hline
}
{Matter exponents for the $c\equ 1/2$, $\t\equ 3$ model.\label{t:3:3}} 

If one repeats this analysis na\"\i vely for $c\equ 4$, 
without taking into account geometric properties, 
no consistent scaling behaviour is found. 
For example, we find Onsager exponents for $\t\equ 1$, but
these change when $\t$ is increased. 
Although the spins in the ``thin" phase cannot be critical,
and contribute little to the space-time volume,
the ``transition'' region, where the spatial length $l$
changes from cut-off length to $l$'s satisfying \rf{3:sca}, 
apparently spoils the measurements, and there are
considerable finite-size effects.
The situation does not improve when the
volume $V_{ext}$ is used instead of the total volume
in the finite-size scaling. 

It seems that the only way to study the critical 
matter behaviour for the case of eight Ising models is to isolate 
explicitly the contributions from
the spins on the extended part of the Lorentzian  geometry.
For this we adopt the following procedure: 
for each configuration we measure the energy $E$ and 
magnetization $M$ on
all vertices belonging to spatial slices whose length is greater 
than a cut-off $l_0(N)=c_{\t} N^{1/\d_h}$, and on all links 
contained in such slices or connecting two of them.
The constants $c_{\t}$ and $\d_h$ are determined from the 
scaling regions of the length distributions 
$SV_N(l)$, with $c_{\t}\approx 0.5$, as discussed in connection 
with eq.\ \rf{3:sca}.  
Denote the numbers of such vertices and links by
$N'$ and $N_L'$. We then compute the averages $e=E/N_L'$ and 
$m=M/N'$, and
measure the expectation values $\vev{N'}$ and $\vev{N_L'}$. 

Looking at the Monte Carlo time histories, 
we observe that whenever $N'\ne 0$, $e$ and $m$ fluctuate
stably around their mean values (even when the vertex number 
$N'$ is close to $0$),
whereas $E$ and $M$ vary slowly but considerably together with
$N'$. We can thus safely ignore the (relatively few) 
configurations with $N'\equ 0$. 
We have also computed the volume $V'\equiv V_{ext}$ 
contributing to the scaling region of $SV(l)$ 
and performed finite-size scaling of the observables
computed from the modified energy and magnetization averages 
$e$ and $m$. 
The results of this final analysis are summarized in
Table~\ref{t:3:4}. We have used a variety of different definitions 
of the system size, to demonstrate that the critical matter
exponents extracted from finite-size scaling do not depend on them.
We conclude that the critical matter behaviour of our model of eight
Ising spins, on the part of space-time that possesses a non-trivial
spatial extension, is governed by the Onsager exponents, and therefore
lies in the same universality class as the model containing only
a single copy of Ising spins.
\TABULAR{|c|c|c|c|c|c|}{
\hline
Observable  & Exponent &  Onsager Value & $V'$-scaling & 
$\vev{N'}$-scaling &  $\vev{N_L'}$-scaling \\
\hline
$\chi$        & $\frac{\g}{\n d_H}$ & 0.875 & 0.85(1)  &0.86(1) & 0.85(1)\\
$D_{\ln |m|}$ & $\frac{1}{\n d_H}$  & 0.5   & 0.520(5) &0.520(2)& 0.48(1)\\
$D_{\ln m^2}$ & $\frac{1}{\n d_H}$  & 0.5   & 0.511(5) &0.512(4)& 0.48(1)\\ 
\hline
}
{Critical matter exponents for the $c\equ 4$, $\t\equ 3$ model. 
We compare scalings with respect to
different definitions of system size.\label{t:3:4}} 

\section{Discussion}\label{discussion}

In order to provide an interpretation for some of our results on 2d
Lorentzian gravity coupled to multiple Ising spins, we first need to
recall some characteristic geometric features of 2d Euclidean gravity.
Consider the one-dimensional spherical ``shell" consisting of all
points separated from a given reference point\footnote{When talking
about ``reference points", we always have in mind averages, calculated
in the statistical ensemble of 2d Euclidean geometries, with each
geometry weighted by the exponential of its classical action.}  by a
geodesic distance $r$.  This curve will in general be multiply
connected.  Let $\rho(l,r)$ denote the number of connected shell
components of length $l$ at distance $r$. It is a remarkable and
universal result in 2d Euclidean quantum gravity that $r$ and $l$ have
a relative anomalous scaling of the form
\beq{d1}
l \propto r^2.
\eeq
For pure 2d Euclidean quantum gravity this was first proved
analytically in \cite{kawai1}, where in the limit of infinite
space-time volume $\rho$ 
was found to be
\beq{d2}
\rho(l,r) \propto \frac{1}{r^2}
\left( c_1 z^{-5/2} + c_2 z^{-1/2} + c_3 z^{1/2}\right) \e^{-z},
~~~z= l/r^2.
\eeq
It was later checked numerically \cite{check,gpk} for various values $c <
1$ of the central charge that in the infinite-volume limit the length
distribution $ r^2 \rho(l,r)$ is only a function of the variable
$z\equ l/r^2$.  In addition, for $z > 1$ the functional dependence on
$c$ turns out to be rather weak.  For a finite space-time volume $N$,
it was found that $r^2 \rho(l,r)$ can be approximated well by
\beq{d2a}
r^2 \rho(l,r) \propto f(z, l/N^{2/d_H}).
\eeq

We can use this relation to calculate the expectation values of integer powers
of the length $l$,\footnote{For $n\equ 1$
eq.\ \rf{d3} is not valid and one obtains instead 
$$\la l \ra_r \propto r^{d_H-1} H(r/N^{1/d_H}),~~~H(0) > 0,$$ where
the Hausdorff dimension $d_H$ is a function of the central charge $c$
of the conformal matter theory coupled to 2d Euclidean quantum
gravity. This contribution comes entirely from small loop lengths $l
\ll r^2$, and is suppressed in the higher moments of $l$.}
\beq{d3}
\la l^n \ra_{r,N} \equiv \sum_l l^n \rho(l,r) \buildrel{N\; {\rm large}}\over
\longrightarrow  N^{2n/d_H} F_n(r/N^{1/d_H}),
~~~~n > 1,
\eeq
where the functions $F_n$ behave like \cite{check}
\beq{d3a}
F_n(x) \sim x^{2n}~~~{\rm for}~~~x < 1.
\eeq
For small $r \ll N^{1/d_H}$ we thus obtain
\beq{d3b}
\la l^n \ra_{r,N} \sim r^{2n}~~~{\rm for}~~~n >1,
\eeq
which is in accordance with relation \rf{d1}, whereas for ``cosmological" 
distances $r \sim N^{1/d_H}$ one finds
\beq{d3c}
\la l^n \ra_{r,N} \sim N^{2n/d_H}~~~~{\rm for}~~~n> 1.
\eeq
It should be emphasized again that eqs.\ \rf{d3a} and \rf{d3b} seem to
be universally true for 2d Euclidean quantum gravity theories with $c
< 1$ and require no cut-off in the continuum limit. To our knowledge
they are the only non-trivial relations in 2d Euclidean quantum
gravity independent of the central charge $c$.

For Lorentzian gravity coupled to a $c\equ 4$ conformal field theory
we saw above that the geometry had undergone a phase transition compared to
$c\equ 0$ and $c\equ 1/2$. In those cases, a continuum limit could
only be obtained if time and space had identical scaling dimensions,
dim$\, l$ = dim$\, t$.  Under the natural identification of Lorentzian
proper time $t$ with the geodesic distance $r$ of the Euclidean
formulae, this should be contrasted with dim$\, l$ = 2\,dim$\, r$,
which follows immediately from relation \rf{d1}. The analogue of
relation \rf{d3c} for Lorentzian gravity with $c\equ 0,1/2$ is given
by
\beq{d4}
\la l^n \ra_{t,N} = N^{n/d_H},~~~~d_H = 2,\hspace{1cm} n>0.
\eeq
More precisely, eq.\ \rf{d4} can be computed exactly for $c\equ 0$ and
is deduced for $c\equ 1/2$ by numerical comparison of the length
distributions.  However, the scaling relation we observed for $c\equ
4$ was not \rf{d4}, but \rf{d3c} (for $n >0$)! The surprising
conclusion is that with increasing central charge $c$ the geometry
undergoes a transition from a state characterized by \rf{d4}, to one
satisfying \rf{d3c}, which is a generic property of {\it Euclidean}
quantum gravity with $c <1$.

What causes this transition as more and more matter is added to the
model?  As discussed in \cite{book,many}, matter has a tendency to
``squeeze off" parts of space-time. In 2d Euclidean quantum gravity
this pinching can take place anywhere and results in an
ever-increasing number of baby universes.  Eventually, for $c > 1$,
the fractal geometry degenerates into branched polymers, which can
simply be viewed as a conglomerate of baby universes of the size of
the cut-off.

In the Lorentzian case by construction no baby universes can be
formed.  The only possible way for matter to squeeze the geometry is
to pinch constant-time slices to their minimal allowed spatial length
$l\equ 1$.  This effect is very obvious in the Monte Carlo simulations
and becomes more pronounced as the central charge is increased.  In
going to $c\equ 4$ the influence of the matter has become so strong
that a genuine phase transition has taken place. Only $t^{2/3}$ of the
$t$ spatial slices (which typically occur together in a single
extended region) have an extension beyond the cut-off scale. 
On the other
hand their average spatial extension behaves like $t^{4/3}$. The
remaining spatial slices have been pinched to the cut-off scale. On
the fraction of slices with a macroscopic extension, one can
then define a scaling limit, which at large distances is
characterized by a Hausdorff dimension $d_H\equ 3$.
Likewise the relative dimensions of space and time are changed from
their na\"\i ve canonical values dim$\, l$ = dim$\, r$, derived from
\rf{d4}, to dim$\; l$ =2 dim$\;r$, dictated by \rf{d3c}.

From our experience with Euclidean quantum gravity, this behaviour may
seem unexpected. In that case, a large influence of the matter on the
geometry is always accompanied by a large back reaction of the geometry
on the matter, in the sense that the critical matter and gravity
exponents always change simultaneously. (An exception to this is the
relation \rf{d3c}, which is valid for {\it all} $c < 1$ and therefore
contains no information about the conformal field theory and its
coupling to geometry.)  The Lorentzian gravity model behaves
differently: the matter strongly affects the geometry (changing bulk
properties like the Hausdorff dimension and the relative scaling
between time and spatial directions), but these apparently drastic
changes are still not sufficient to alter any of the universal matter
properties. Even when 8 Ising models are coupled to Lorentzian
gravity, the critical matter exponents still retain their Onsager
values.

This situation provides further support for the viewpoint advanced in
our previous work \cite{al,aal} that the critical gravity and matter
behaviour of the Euclidean models is entirely determined by the
presence of baby universes. In the light of our new results, the
argument may be put as follows. So far it has been unclear whether the
change in the critical exponents of conformal field theories when
coupled to Euclidean quantum gravity was due to the strong
back reaction of the geometry on the matter or to the baby universes
that were present {\it a priori}. Lorentzian gravity with 8 Ising
spins provides an example where undeniably the interaction of the
matter and gravity sectors is strong. Nevertheless the critical Ising
exponents remain unchanged. This strongly suggests that in the
Euclidean case it is really the baby universes which are responsible
for the observed chan\-ges in the universal properties of the matter.

While we have not undertaken a systematic search for the exact value
of $c$ where the phase transition in geometry takes place, it is
tempting to conjecture that it occurs at $c_{\it crit} \equ 1$.
Independent of the exact value of $c_{\it crit}$, we have identified a
weak analogue of the $c\equ 1$ barrier also in Lorentzian gravity.
From the point of view of matter, nothing dramatic happens when the
barrier is crossed. However, the behaviour of the quantum geometry
undergoes a qualitative change and even shares some features with the
non-singular part of the quantum geometries described by 2d Euclidean
gravity coupled to matter with $c < 1$.  This highlights the universal
nature of the relation $l \propto r^2$ and motivates the search for a
simple underlying explanation, which may have a status similar to that
of $d_H \equ 2$ for Brownian motions.

\section{Outlook}\label{outlook}

In closing, let us step back to examine the potential larger
implications of all we have learned so far about two-dimensional
Lorentzian quantum gravity. 
Our original aim was to find a non-perturbative path-integral
formulation for quantum gravity. Previous attempts in this
direction had largely been confined to the  sector of
{\it Euclidean} space-time metrics. Since for general metrics
there is no straightforward analogue of the Wick rotation, 
a path integral over Lorentzian geometries is
likely to require a more radical modification (compared with
the Euclidean theory) than a mere analytic continuation of the
action. 

We chose to make the path integral Lorentzian by requiring {\it each
individual} space-time geometry contributing to the state sum
to carry a causal structure associated with a Lorentzian geometry.
In order to make the construction well defined, a regularization
is necessary, and we used the method of dynamical
triangulations (where geometric manifolds are represented
as gluings of $d$-dimensional simplices), which had previously 
been employed successfully in a Euclidean context. An ideal
testing ground for such a proposal is gravity in dimension
$d\equ 2$, whose Euclidean sector (``Liouville gravity") has been 
studied extensively by a variety of methods. We performed the
Lorentzian state sum exactly, 
over a set of dynamically triangulated
2-geometries satisfying a (discrete analogue) of causality, and
taking a continuum limit.  

Maybe surprisingly, the resulting continuum theory turned
out to be a new, {\it bona fide} theory of 2d quantum gravity
fundamentally different from (Euclidean) Liouville gravity.
As already mentioned in the introduction, it
describes an ensemble of strongly fluctuating geometries,
with local curvature degrees of freedom. 
Nevertheless, the geometry is less fractal than its
Euclidean counterpart, and therefore closer to our intuitive, 
classical notions of smooth geometry. 

The existence of at least two different 
quantum gravities constructed by rigorous path-integral methods
in 2d raises the question of ``which one is the {\it right} theory?" 
There is no ultimate answer to this, since two-dimensional
gravity (never mind its signature) does not describe any phenomena
of the real world. Aficionados of Liouville gravity might object
by saying that the Lorentzian version of quantum geometry was surely 
the less interesting, with not enough ``happening" compared with
the Euclidean case where, for example, 
the Hausdorff dimension $d_H$ changes with the matter content.
However, even if this were the case, it would not
disqualify Lorentzian gravity from being a good candidate for a
quantum gravity theory, since  
we do not know what the geometry of ``real'' quantum gravity
looks like at the Planck scale. 
So far there is little evidence to suggest non-smoothness of
the space-time geometry up to the grand unified scale, 
which itself after all is only a few orders
of magnitude removed from the Planck scale. 

At any rate, our present investigation shows that also 
in two-dimensional Lorent\-z\-ian gravity things ``do happen''. 
There is a strong interaction if
one couples a sufficient amount of matter to Lorentzian geometry,
affecting the universal properties of the combined system.
In fact, one can argue that the resulting structure of quantum 
geometry is {\it richer} than that of the corresponding
Euclidean model with matter, since its fractal structure (measured
by the Hausdorff dimension) acquires a scale-dependence.
Moreover, the interaction in Lorentzian gravity is strong, but --
unlike in Euclidean gravity -- not too strong in the sense of
leading to a complete degeneration of the carrier geometry. 

In a similar vein, evidence is accumulating that the structure of
Euclidean quantum gravity with and without matter is governed 
entirely by the presence of baby universes (branching configurations
not present in the Lorentzian version because of their
incompatibility with causality). There is nothing wrong with this:
statistical mechanical models of Euclidean gravity provide
examples of generally covariant systems which are highly interesting 
in their own right. What they might 
teach us about quantum gravity proper is much less clear, since 
related (and from a classical point of view highly degenerate) 
branched-polymer configurations and their associated fractal structure
play a central role in Euclidean gravity in higher dimensions too.
There they seem to affect the theory in an undesirable way, 
making it difficult to obtain an interesting 
continuum limit of the statistical models of Euclidean 
quantum gravity in dimension $d > 2$.

There is then a conclusion to
be drawn for our eventual goal, the quantization of the physical
theory of gravity in four space-time dimensions, whose character
we know to be Lorentzian. Judging from our experience in two
dimensions, the Lorentzian and Euclidean theories ({\it if} they
both exist and are unique) may not be as closely related as is 
sometimes hoped for, invoking the example of 
standard, non-generally covariant quantum field theory. 
Our research has highlighted the importance of imposing 
causality (and thereby suppressing spatial topology changes) in the
path integral over geometries. It remains to be seen which effect
an analogous prescription has for quantum gravity theories in
higher dimensions.

\acknowledgments
J.A. and K.A. thank MaPhySto, Centre for Mathematical Physics and
Stochastics, funded by a grant from The Danish National Research
Foundation, for financial support.
We acknowledge the use of Geomview \cite{geom} as our underlying 3D
geometry viewing program.



\end{document}